\documentclass{iau}

\usepackage{amsmath}
\usepackage{graphicx}
\usepackage{multirow}

\newcommand{\kms}{\mbox{km·s$^{-1}$}}
\newcommand{\seca}{\mbox{\rlap{.}$''$}}
\newcommand{\secs}{\mbox{\rlap{.}$^{\rm s}$}}

\begin{document}

\lefttitle{J.\,Alcolea,\,V.\,Bujarrabal,\,A.\,Castro-Carrizo, {\it et al.}}
\righttitle{The Red Rectangle: a thin disk with big grains}

\journaltitle{Planetary Nebulae: a Universal Toolbox in the Era of Precision Astrophysics}
\jnlDoiYr{2023}
\doival{10.1017/xxxxx}
\volno{384}

\aopheadtitle{Proceedings IAU Symposium}
\editors{O. De Marco, A. Zijlstra, R. Szczerba, eds.}
 
\title{The Red Rectangle: a thin disk with big grains
}

\author{Javier Alcolea$^1$, Valentín Bujarrabal$^1$, 
	Arancha Castro-Carrizo$^2$, \\ Jacques Kluska$^3$, 
	Carmen Sánchez Contreras$^4$, and Hans van Winckel$^3$}
\affiliation{$^1$ Observatorio Astronómico Nacional (IGN/CNIG), MITMA, Spain; \\
	$^2$ Institut de Radioastronomie Millimétrique,  France; 
	$^3$ Instituut voor Sterrenkunde, KU Leuven, Belgium;
	$^4$ Centro de Astrobiología (CAB), CSIC-INTA, Spain}

\begin{abstract}
The Red Rectangle is a nebula surrounding the post-AGB star HD 44179. It is the prototype of a particular class of nebulae
associated with post-AGB binaries characterised by the presence of stable circumbinary disks in (quasi-)Keplerian
rotation. Here we present the results of new high-resolution (0{\mbox{\rlap{.}$''$}}02 -- 0{\mbox{\rlap{.}$''$}}05) ALMA observations of continuum
and line emissions at 0.9\,mm. The continuum maps are analysed through a simple model of dust emission,
which can reproduce the observational data. We find that most dust emission in the Red Rectangle is concentrated in the central
regions of the rotating disk and that the settlement of dust grains onto the equatorial plane is very significant, particularly in
comparison with the much larger scale height displayed by the gas distribution. The diameter of the dust-emitting region is
about 250\,au, with a total width of about 50\,au. This region coincides with the warm PDR where certain molecules (like HCN),
CI, and CII are presumably formed, as well as probably PAHs. From the spectral index, we confirm the presence in the disk of large
grains, with a typical radius of about 150\,$\mu$m, which supports the long-lived hypothesis for this structure. We
also confirm the existence of a compact ionised wind at the centre of the nebula, probably emerging from the accretion disk around the 
companion, for which we derive an extent of about 10\,au and a total flux of 8\,mJy. We also briefly present the results on molecular
lines of $^{12}$CO, $^{13}$CO, and other less abundant species.
\end{abstract}


\maketitle

\section{Introduction}

There is a class of binary stars consisting of a post-AGB primary and a 
main-sequence secondary, with about 85 members in our galaxy \citep{vanwinckelARAA2003,kluska2022}. These sources are systematically surrounded by circumbinary disks in (sub-)Keplerian rotation \citep{deroo2006, valentin2013a, valentin2013b, valentin2015, valentin2016, kluska2019}. Very often, they also present a disk-wind bipolar outflow \citep{ivan2022,ivan2023}. It is also believed that these disks are responsible for the observed orbital periods of $\sim$\,150\,--\,2000\,d: these values lie in between the bimodal distribution of periods predicted for post-AGB binaries \citep{nie2012,oomen2018,vanwinckel2018}, suggesting a strong angular momentum exchange between the stellar system and the circumbinary rotating disk.

One of the best-studied members of this class is the post-AGB star HD 44179 in Monoceros, which lies at the centre of an X-shaped nebula known as the Red Rectangle \citep{cohen2004,valentin2013b,valentin2016}. At a distance of 710 pc, this source has been modelled in detail by \citet{menchi2002} from IR imaging. The axis of the nebula is slightly off the plane of the sky, and the equatorial plane is oriented at a parallactic angle (PA) of 103$^\circ$. The binary has a period of 317\,d, and the total stellar mass is 1.8\,M$_\odot$ \citep{thomas2011}. The source has a peculiar chemistry with an O-rich component  and strong C-rich extended red emission (ERE) and PAH IR features in the nebula \citep[see {\em e.g.}][]{cohen2004}, which in addition to the detection of CI, CII, and some molecular lines, points to the existence of a photon-dominated region (PDR) at the centre of this object \citep{valentin2016}.

The nebula was previously mapped with ALMA at a resolution of 0{\mbox{\rlap{.}$''$}}5 by \citet{valentin2013b,valentin2016}, revealing the structure and velocity field of the molecular gas in both the disk and the biconical outflow. Here we present new band 7 ALMA observations of the Red Rectangle but at a much higher spatial resolution, down to 0{\mbox{\rlap{.}$''$}}020 (20\,mas). 

\section{The new ALMA observations}

The observations presented here have been obtained at the ALMA project 2019.1.00177.S \citep{valentin2023}. These are band 7 observations in two array configurations with baseline lengths from 14\,m to 14.9\,km. The observed frequency ranges and native spectral resolutions, as well as the main properties of the spectral lines detected in addition to the continuum, are given in Table \ref{tableobs}.

\begin{table}[h!]
\centering
\caption{Observed continuum ranges and detected spectral lines}
{\tablefont
\begin{tabular}{@{\extracolsep{\fill}}cccrc}
\midrule
Observed frequency range and  & Detected lines & Adopted rest & \multicolumn{1}{c}{Upper level} & Spectral resolution\\
native spectral resolution    &                & frequency    & \multicolumn{1}{c}{energy}      & used in the maps\\
\midrule
330.350 -- 330.827\,GHz (0.21\,\kms) & $^{13}$CO $J$=3--2                                  & 330\,587.965\,MHz &   31.7\,K & (0.5\,\kms) \\
\midrule
330.744 -- 332.658\,GHz (0.85\,\kms) & H$_2$O $\nu_2$=2 $J_{K_a,K_c}$=3$_{2,1}$--4$_{1,4}$ & 331\,123.730\,MHz & 4881.4\,K & (1.0\,\kms) \\
\midrule
343.563 -- 345.448\,GHz (0.85\,\kms) & SiO $\nu$=1 $J$=8--7                                & 344\,916.247\,MHz & 1843.6\,K & (1.0\,\kms) \\
                                     & H$^{13}$CN $\nu_2$=1 $J$=4--3 $l$=1e                & 345\,238.771\,MHz & 1057.2\,K & (1.0\,\kms) \\
                                     & H$^{13}$CN  $J$=4--3                                & 345\,339.769\,MHz &   41.4\,K & (1.0\,\kms) \\
\midrule
345.558 -- 346.032\,GHz (0.21\,\kms) & $^{12}$CO $J$=3--2                                  & 345\,795.990\,MHz &   33.2\,K & (0.5\,\kms) \\
\midrule
\end{tabular}
}
\label{tableobs}
\end{table}

The attained spatial resolution is about 20 mas (14\,au) for the continuum data (using robust weighting) and about 55 mas (40\,au) for the spectral line data (using natural weighting). All data has been calibrated with the standard ALMA pipeline plus self-calibration using the pure continuum emission. We have verified that the present data are consistent with the previous 0{\mbox{\rlap{.}$''$}}5 resolution ALMA observations, and that no flux is filtered out in the new dataset within the 20\% calibration uncertainty. Additional information on the observations and data analysis procedure can be found in \citet{valentin2023}.

\section{Continuum emission results} \label{continuum}

\begin{figure}
\begin{center}
 \includegraphics[width=0.9\textwidth]{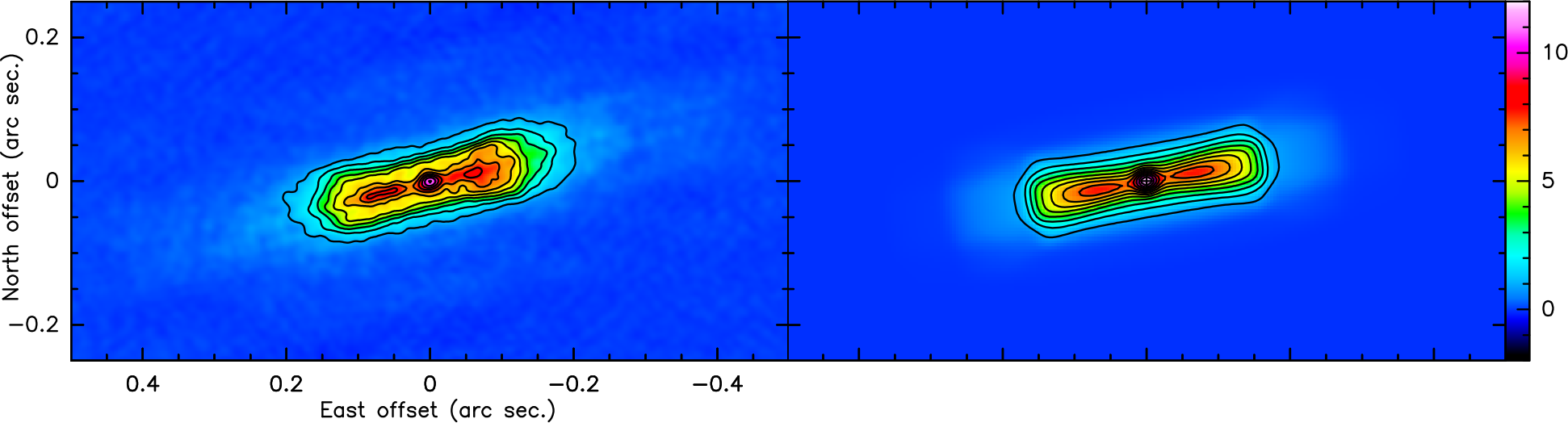}
\end{center}
\caption{To the left, the new ALMA 0.9\,mm continuum map of the Red Rectangle. The size of the circular restoring beam is 20 mas (14\,au at 710 pc).
To the right, results from our best-fit model of the continuum emission of the Red Rectangle at 0.9 mm as observed by ALMA (see Sect.\,\ref{modelling}). The model includes the contribution of the dust grain emission (DDC) and of the free-free of the central ionised wind (CCC).   Intensity units are mJy·beam$^{-1}$. Colour-coded scale is on the right side. Plotted contours are 1 to 12 by 1\,mJy·beam$^{-1}$ 
(36 to 432 by 36\,K in Rayleigh-Jeans equivalent brightness temperature). Colour-code scale and contours are the same in both real observed data and the results of the radiative transfer code.
}
\label{figcont}
\end{figure}

The image of the continuum emission, shown in Fig. \ref{figcont}, left panel, consists of a central compact component (CCC) plus the contribution of a thin disk seen almost edge-on (dust disk component, DDC hereafter). The orientation of the equatorial plane is at PA=103$^\circ$, with the symmetry axis close to the plane of the sky. For the central peak, we derive J2000 coordinates of R.A.=06$^{\rm h}$19$^{\rm m}$58\secs1988 and Dec.=--10$^\circ$38$'$15\seca216, which have been adopted as the centre of our maps. Comparing these coordinates with those from previous VLA and ALMA observations \citep{jura1997,valentin2013b,valentin2016}, we derive proper motion values of $\mu_x$=--13.4$\pm$0.1\,mas·a$^{-1}$ and $\mu_y$=--26.3$\pm$1.4\,mas·a$^{-1}$. 

The total detected fluxes are 664 and 722 mJy at 331.7 and 344.5 GHz respectively, from which we estimate a spectral index (SI) of about 2, compatible with the emission from dust grains of relatively large size \citep[see][]{jura1997}. To isolate the contribution of the CCC we have performed a compact source model fitting in the uv-plane just considering baselines longer than 5\,km, eliminating the contribution from the more extended structures of the DDC. We have obtained a total flux of 8.3\,mJy and a size of 15 mas (10\,au) for the CCC, which we attribute to the free-free emission of an ionised gas wind \citep[see][]{jura1997}. After removing this contribution of the CCC, we find that the strongest continuum emission comes from a region 200 mas wide in diameter (140\,au) and 25 mas in height (18\,au).

We can get a better estimate of the SI for both CCC and DDC by comparing these new fluxes with those from the literature at cm- and sub-mm-wavelengths \citep{jura1997,valentin2013b,valentin2016}. The results of the SIs fittings are shown in Fig.\,\ref{figslopes}, left panel. The derived SI for the CCC is 0.8, very close to the expected value of 0.6 for an isothermal ionised wind of constant mass loss \citep{panagia}. This confirms the free-free nature of this emission but suggests deviations from the simple case of a homogeneous wind with a $1/r^2$ density. Using the CCC flux value at 0.9\,mm, and assuming that the size measured for this component is several times the ``so-called'' characteristic radius, we derive a temperature of 3\,600 -- 10\,000\,K and a mass loss rate of 8\,10$^{-8}$ -- 1.7\,10$^{-7}$\,M$_\odot$·a$^{-1}$ for this ionised wind, very probably released by an accretion disk around the companion \citep{witt2009}. The derived SI for the DDC, 2.13, is much closer to the value expected for an optically thick emission or the emission by very big grains regardless of the opacity (SI\,=\,2.0), than to the value expected for an optically thin emission of small grains (SI\,=\,3). In addition, the brightness measured in the disk reaches values larger than 200 K, implying that the opacity can not be very small (because the temperature of the grains cannot exceed $\sim$\,1000 -- 1500\,K). This suggests the presence of effective grain growth \citep[see][]{jura1995,jura1997} and moderate opacities in the strongest emitting regions of our disk. 

\section{Continuum modelling} \label{modelling}

\begin{figure}
\begin{center}
\begin{minipage}[b]{0.3\textwidth}
    \includegraphics[width=\textwidth, height=3.75cm]{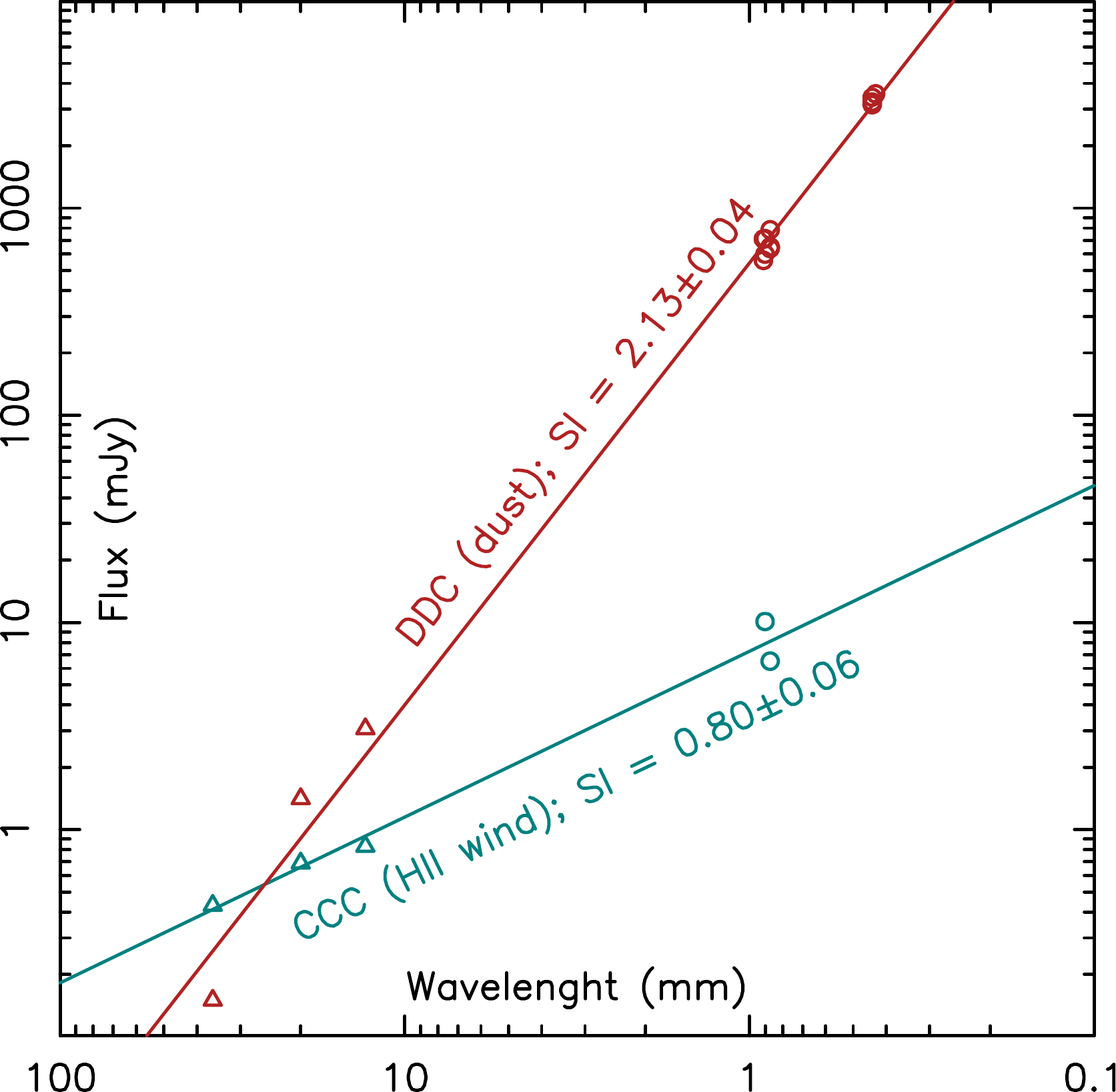}
\end{minipage}
\begin{minipage}[b]{0.6\textwidth}
     \includegraphics[width=\textwidth]{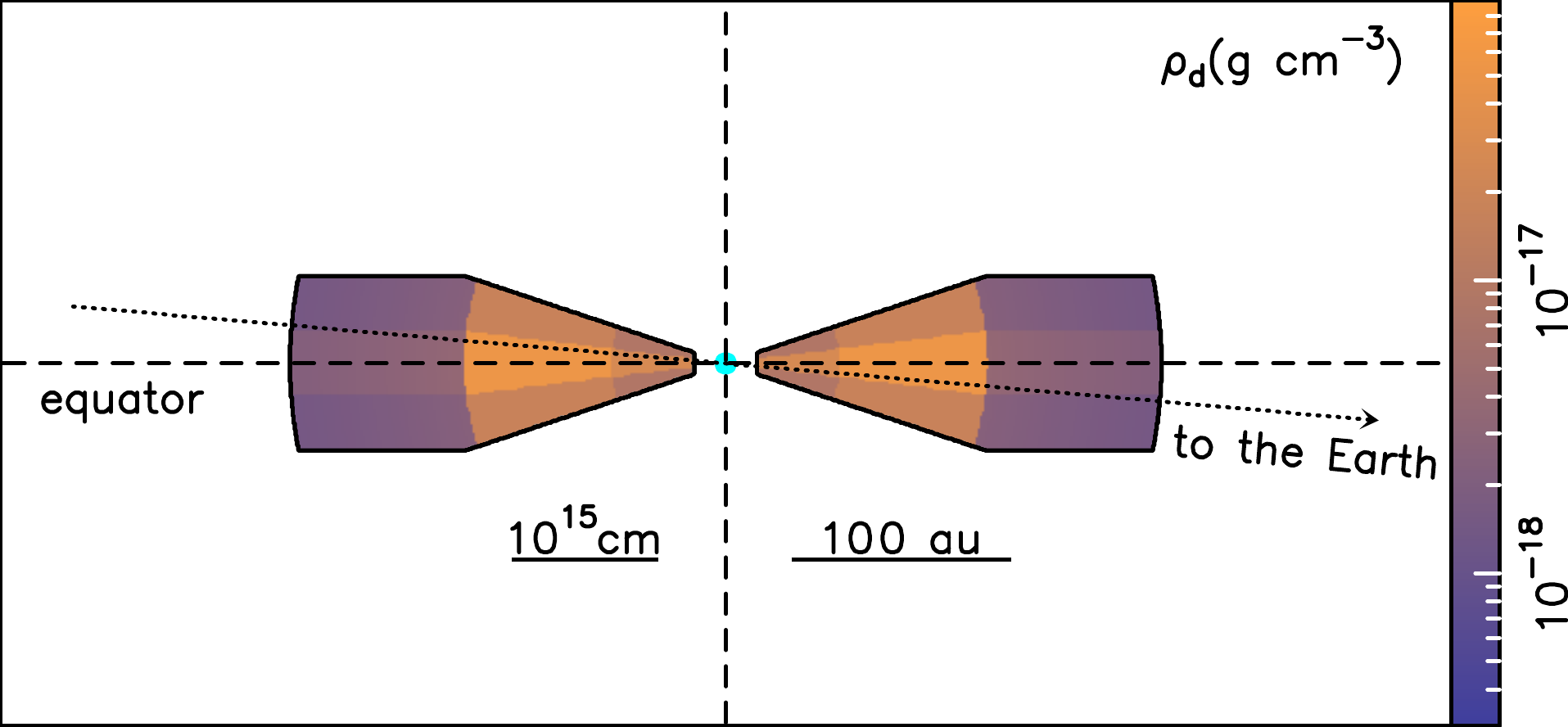}
\end{minipage}
\caption{To the left, total integrated continuum fluxes in the Red Rectangle for the CCC in cyan, and the more extended DDC in red: we are assuming that the DDC flux is the total flux minus that from the CCC. cm-wave data (triangles) are from \citet{jura1997}, while sub-mm measurements (circles) are from \citet{valentin2013b,valentin2016}, and this work \citep[see][]{valentin2023}. The spectral indexes (and the corresponding formal errors), computed from regression lines, are also indicated. To the right, a sketch of the geometry, sizes, density distribution (grey scale), and viewing angle used for the model of the best fit of the data. The cyan dot at the centre represents the CCC, the compact ionised wind discussed in Sect.\,\ref{continuum}. See Sect.\,\ref{modelling} and \citet{valentin2023} for additional details.
}
\label{figslopes}
\end{center}
\end{figure}

To derive the physical parameters of the dusty disk that reproduces the continuum emission of the Red Rectangle at 0.9\,mm, we have developed a simple model of radiative transfer for a dusty disk. Following \citet{tielens}, we have adopted very simple laws for grain absorption/emissivity (scattering is not important at these wavelengths). We have assumed constant emissivity for wavelengths smaller than a characteristic one, about 10 times the grain radius, $r_g$, and an emissivity that decreases inversely proportional to the wavelength elsewhere. We have also adopted a grain density of 3.0\,gr·cm$^{-3}$, typical of O-rich dust material. The best fitting of the data has been obtained for a $r_g$ value of 150\,$\mu$m, comparable to the value of 200\,$\mu$m previously inferred by \citet{jura1997}. The density structure of the disk is described in Fig.\,\ref{figslopes}, right panel. The temperature of the dust varies inversely to the square root of the distance to the centre, with a value of 400\,K at 1.8\,10$^{15}$\,cm (120\,au), and an upper limit of 1200\,K. The predicted emission from the model, directly comparable to the observations, is shown in Fig.\,\ref{figcont}, right panel, where we have also included the contribution from a small central component to account for the free-free emission of the CCC. Our best model geometry and density distribution are compatible with those of \citet{menchi2002}. For the assumed grain density, we derive a total dust mass in the disk of 5\,10$^{-5}$ -- 10$^{-4}$ M$_\odot$. This is 150 times smaller than the total molecular mass of the disk \citep{ivan2023}. This gas/dust mass ratio value may look a bit low for O-rich environments but note that these circumbinary disks around binary post-AGB sources, including the Red Rectangle, are molecule deficient compared to standard AGB circumstellar envelopes \citep[see][]{ivan2022}. 

Apart from effective grain growth, our model also finds a strong settlement of solid material onto the equatorial plane of the disk: 80\% of the disk's solid mass lies at distances from the equatorial plane smaller than 5\,10$^{14}$\,cm ($\sim$\,35\,au). The extent of this dense disk is also relatively small, with a width-to-height ratio of about 5: disk radius of 2.5\,10$^{15}$\,cm ($\sim$\,175\,au). These sizes are significantly smaller than the distribution shown by the molecular gas detected in quasi-Keplerian rotation \citep[see][and Sect. \ref{sectlines}]{valentin2016}. The temperature of this dense dust disk varies between 350 and 1200\,K at its outer and inner layers, respectively.

\section{Molecular line results} \label{sectlines}

\begin{figure}
\begin{center}

    \includegraphics[width=0.9\textwidth]{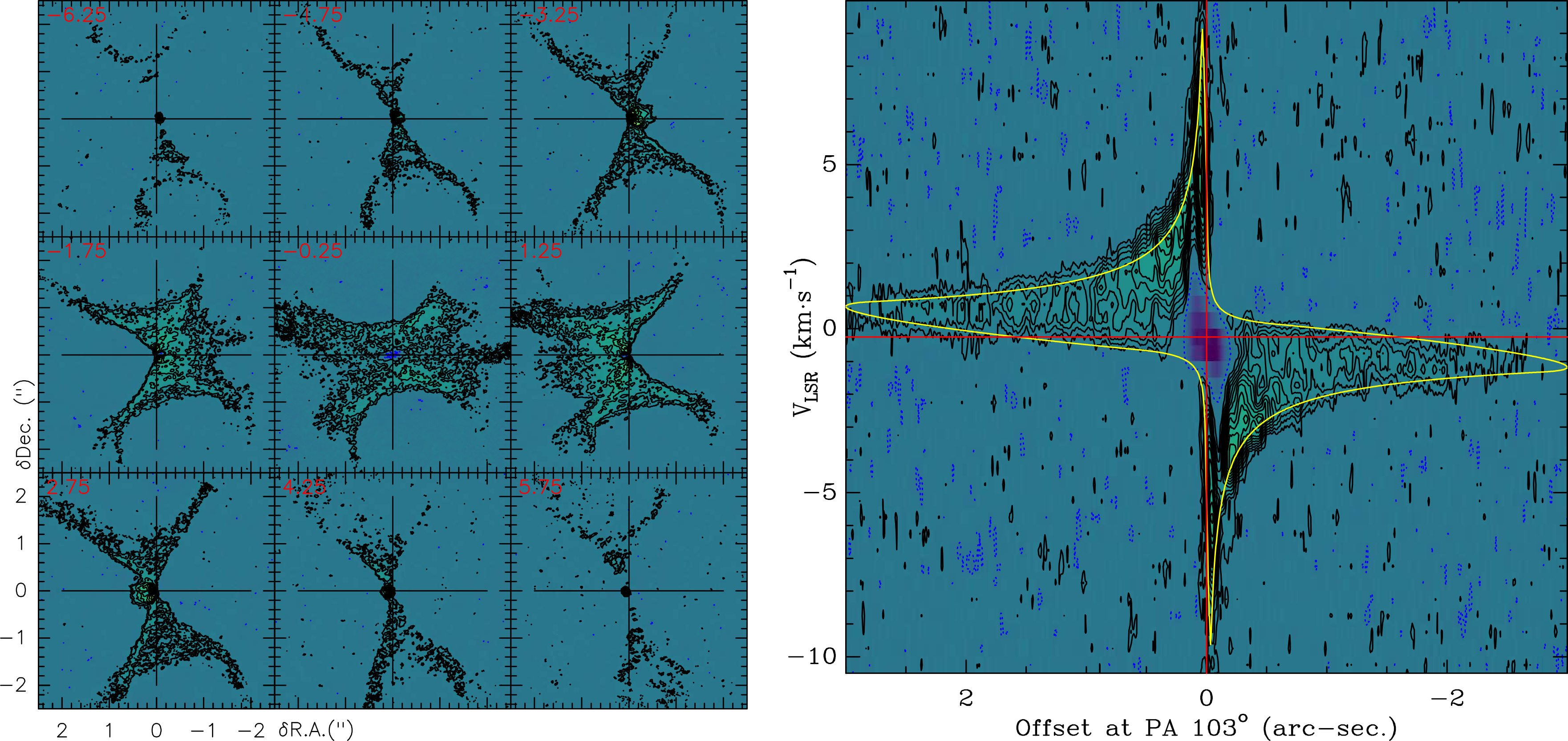}

\caption{Results from the new ALMA band 7 observations of the Red Rectangle for the $J$=3--2 line of $^{12}$CO. Here the restoring beam is $\sim$\,55\,mas and the velocity resolution is 0.5\,\kms. On the left panel, maps of the $^{12}$CO emission for selected velocities (indicated in red figures at the top right corner). To the right, the velocity-position (PV) diagram along the equatorial direction (at PA=103$^\circ$, i.e., positive offsets are in the east by south-east direction). The red cross marks the assumed central position and systemic velocity of the source (--0.25\,\kms). The yellow line comprises the locations of the gas in the PV diagram according to the velocity law described in Sect.\,\ref{sectlines}. In both panels, the contours are drawn every 0.005\,mJy·beam$^{-1}$ ($\sim$\,20\,K); negative levels are drawn in dashed blue.}
\label{12co32}
\end{center}
\end{figure}

\begin{figure}
\begin{center}

    \includegraphics[width=0.9\textwidth]{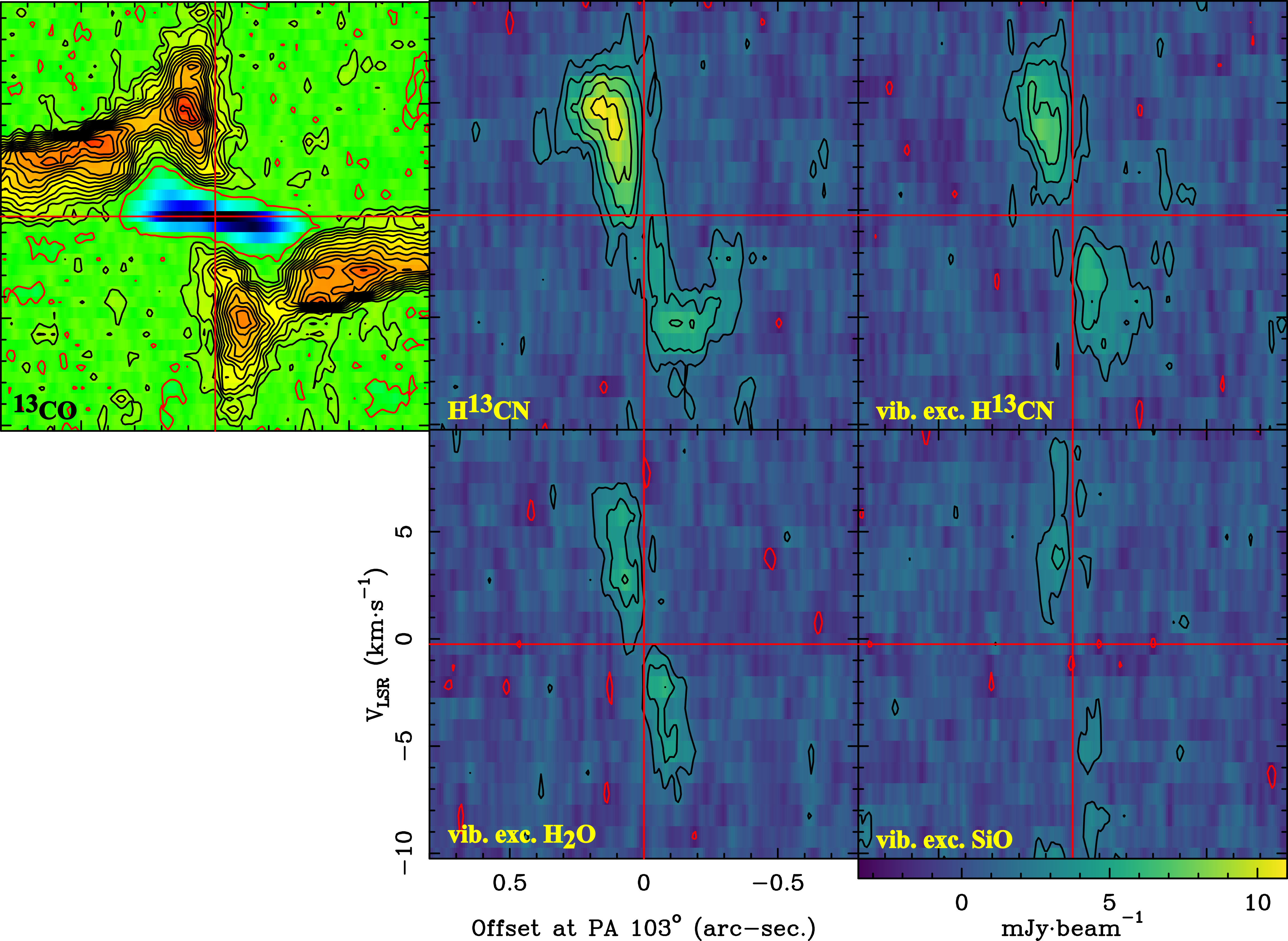}

\caption{Results from the new ALMA band 7 observations of the Red Rectangle for species other than $^{12}$CO. Velocity-position (PV) diagrams along the equatorial direction (PA=103$^\circ$) for $^{13}$CO (top left), ground and vibrationally excited H$^{13}$CN (top middle and top right respectively), and for vibrationally excited H$_2$O and SiO (bottom middle and bottom right respectively). The restoring beam is $\sim$\,55\,mas and the velocity resolution is 0.5\,\kms\ for $^{13}$CO and 1.0\,\kms\ in the rest of the cases. The red cross marks the assumed central position and systemic velocity of the source (--0.25\,\kms). The plot zooms into the central region of the disk where species other than CO are detected. On the top left panel, $^{13}$CO clearly shows the signature of the quasi-Keplerian rotation. Here positive/negative contours are drawn in black/red. The negative values at the centre of the $^{13}$CO panel (cyan and dark blue pixels) are due to the subtraction of the continuum emission and the presence of cold molecular gas in front of the dust disk. The rotation signature is barely hinted at in the H$^{13}$CN emission, which extends about $\pm$0{\mbox{\rlap{.}$''$}}35. For the vibrationally excited emissions, just the inner rim of the molecule-rich rotating disk, extending just $\pm$0{\mbox{\rlap{.}$''$}}15, is detected. Except for $^{13}$CO, contours are drawn every 0.0025\,mJy·beam$^{-1}$ (equivalent to $\sim$\,10\,K) and negative levels are drawn in dashed blue: the colour-code scale is shown below the SiO emission panel.}
\label{cortes}
\end{center}
\end{figure}

Although the flux filtered out in these new high-resolution observations is low or negligible, even for the most sensitive molecular gas probes like $^{12}$CO, the new observations mostly trace the rotating disk; the X-shaped wind is detected very noisily due to sensitivity limitations (see Fig.\,\ref{12co32}, left panel). On the contrary, the disk and its characteristic Keplerian-like rotation are traced with superb detail in both CO lines (see Figs.\,\ref{12co32}, right panel, and \ref{cortes} respectively for $^{12}$CO and $^{13}$CO). The size of the molecular disk is much larger than that traced by the dust, 6$''\times$1$''$ ($\sim$\,4000$\times$700\,UA) in $^{12}$CO (2{\mbox{\rlap{.}$''$}}0$\times$0{\mbox{\rlap{.}$''$}}5 in the case of $^{13}$CO), i.e. up to $\sim$\,10 times bigger. We do not see signs of gas concentration or depletion at the equatorial plane: the negative flux seen at low velocities at the nebula's centre is due to the removal of the continuum emission from the line maps and the presence of opaque cold gas in between the continuum source and the observer. In the disk, $^{12}$CO is mostly opaque: thus, the  observed intensities provide direct information on the excitation of the gas. We derive $T_{\rm exc}$ ($\approx$ $T_{\rm kin}$ for CO) up to 300\,K, similar to the dust temperatures. $^{13}$CO is partially opaque in the disk: therefore the estimation of parameters such as the $^{12}$C/$^{13}$C ratio and the molecular disk mass requires detailed modelling. In the disk, we detect an inner radius of the molecular gas emission, $R_{\rm int}$, of $\sim$\,0{\mbox{\rlap{.}$''$}}04 (28\,au): we are not claiming that there is a central region devoid of molecular gas, just that these central parts do not emit in any of the detected lines above our detection threshold (of about 10\,K). Reconciling the observed velocities with central stellar mass requires the presence of rotation and radial expansion that decrease with the square root of the distance to the center: $V_{\rm rot}  = 8 \sqrt{(R_{\rm int}/r)}$\,\kms\ and $V_{\rm exp}  = 5 \sqrt{(R_{\rm int}/r)}$\,\kms\ (see Fig.\,\ref{12co32}, right panel).

In addition to the previously mapped lines of $^{12}$CO and $^{13}$CO $J$=3--2, and of H$^{13}$CN $J$=4--3 \citep{valentin2013b,valentin2016}, we have also detected vibrationally excited lines of H$^{13}$CN, H$_2$O and SiO, with upper-level excitation energies over 1000\,K (see Table\,\ref{tableobs}). The emission of these three vibrationally excited lines is very similar (see Fig.\,\ref{cortes}): they show a very compact central emission, just tracing the inner regions of the rotating disk, from a very narrow range of distances from the centre, $\sim$\,150\,mas (100\,au). This is probably related to the presence of the central PDR \citep{valentin2016}, and/or strong IR-pumping due to the dense warm dust at these locations. In the case of H$^{13}$CN $J$=4--3 from the ground vibrationally state, we observe a slightly more extended emission, up to 250\,au from the centre. The detection of both C-rich (H$^{13}$CN) and  O-rich species (H$_2$O), together with the detection of CI, CII, and PAHs strongly support the existence of a PDR at the centre of the nebula \citep{valentin2016}. ALMA mapping of CI is mandatory to confirm the presence of this PDR and derive its main properties.
	
\begin{acknowledgements}
This work is supported by the I+D+i coordinated project Nebulae Web, PID2019-105203GB-C21 EVENTs (JA, VB, ACC) and PID2019-105203GB-C22 Genesis (CSC), funded by the Spanish MCIN/AEI grant 10.13039/501100011033. The results presented in this contribution were obtained at the ALMA project 2019.1.00177.S (PI V. Bujarrbal). ALMA is a partnership of ESO (representing its member states), NSF (USA) and NINS (Japan), together with NRC (Canada) and NSC and ASIAA (Taiwan), in cooperation with the Republic of Chile. The Joint ALMA Observatory is operated by ESO, AUI/NRAO and NAOJ.
\end{acknowledgements}


\end{document}